# Large tunable intrinsic gap in rhombohedral-stacked tetralayer graphene

K. Myhro[1‡], S. Che[1,2‡], Y. Shi[1], Y. Lee[1], K. Thilahar[1], K. Bleich[1], Dmitry Smirnov[3], C. N. Lau[1,2*]

[1]Department of Physics and Astronomy, University of California, Riverside, Riverside, CA 92521
[2]Department of Physics, The Ohio State University, Columbus, OH 43210
[3]National High Magnetic Field Laboratory, Tallahassee, FL 32310

[‡]These authors contributed equally to this work

ABSTRACT
**In rhombohedral-stacked few-layer graphene, the very flat energy bands near the charge neutrality point are unstable to electronic interactions, giving rise to states with spontaneous broken symmetries. Using transport measurements on suspended rhombohedral-stacked tetralayer graphene, we observe an insulating ground state with a large interaction-induced gap up to 80 meV. This gapped state can be enhanced by a perpendicular magnetic field, and suppressed by an interlayer potential, carrier density, or a critical temperature of ~ 40 K.**

Since 2004, graphene has rapidly become an extra-ordinary 2D electron system for low dimensional physics, as it hosts massless Dirac fermions exhibiting an unconventional quantum Hall effect with a with a Berry's phase of $\pi$[1,2]. More recently, the few-layer "cousins" of monolayer graphene (MLG) have also attracted significant attention, as they constitute highly unusual, fascinating 2D platforms[3-6]. Like MLG, they are atomically thin membranes with chiral charge carriers; however, they differ from MLG in band structure, crystal symmetries, and strength of electronic interactions, all of which have profound effects on their electronic properties[7-10].

Few-layer graphene (FLG) has two natural stable allotropes which can be distinguished by Raman spectroscopy[11,12]. ABA or Bernal stacking is the most stable and abundant form found in 85% of natural graphite. ABC or rhombohedral stacking, which occurs in ~14% of bulk graphite, obeys inversion symmetry, and its dispersion can be approximated as simply $E \sim k^M$, where $k$ is the wave vector and $M$ the number of layers[13]. Thus, rhombohedral-stacked FLG (r-FLG) is highly unusual in the very flat bands near the charge neutrality, which host large and even diverging (for $M>2$) density of states and extremely large electronic interactions. The interaction parameter, $r_s$, also known as the Wigner-Seitz radius, is the ratio of the average electron Coulomb interaction energy to the Fermi energy, given by $r_s \propto n^{-(M-1)/2}$, where $n$ is charge density and $M$ is the power of the dispersion relation. For MLG, $M=1$, $r_s = e^2/\varepsilon_r \hbar v_F \sim 2.2$ in suspended MLG, independent of density. Here, $e$ is the electron charge, $\varepsilon_r$ is the background dielectric constant, ~1 for suspended graphene, $\hbar$ is reduced Planck's constant and $v_F \sim 10^6$ m/s is the Fermi velocity. For bilayer graphene (BLG) and rhombohedral-stacked trilayer (r-TLG) and tetralayer (r-4LG) graphene, $r_s \propto n^{-1/2}$, $n^{-1}$, and $n^{-3/2}$, respectively. Close to the charge neutrality point, $r_s$ increases by 1-2 orders of magnitude when an extra layer is added. Indeed, interaction-induced gaps of ~ 2 meV in suspended BLG[14-17], and ~42 meV in suspended r-TLG[18, 19] have been observed (see Table 1). Among theoretical proposals[20-36], the experimental observations are most consistent with a layer antiferromagnetic state (LAF) with broken time reversal symmetry.

---

[*] Email: lau.232@osu.edu

A natural question is whether the trend of growing magnitude of the interaction-induced gap with increasing layer thickness persists for $M>3$. On the one hand, thicker r-FLG has even larger density of states at the charge neutrality point that are ever more unstable against electronic interactions, thus favoring the formation of a larger gap; on the other hand, screening in thicker FLG is stronger, thus reducing the interaction strength, while charge orbitals become less confined to the atomic plane. Interplay between these factors suggests that electronic interactions should reach a maximum at certain critical thickness and decrease for thicker r-FLG. In fact, a theoretical calculation predicts that the gap is the largest in r-TLG and decrease monotonically for $M>3$[37]. However, experimentally studies of thicker r-FLG have been absent to date.

Here we report that, continuing the trend of increasing interaction in thicker r-FLG, we observe an insulating state in r-4LG, with a gap up to 80 meV. The energy gap increases further with a perpendicular magnetic field, but closes upon the application of an out-of-plane electric field of either polarity, increasing charge density or raising temperature. Our results are consistent with a LAF state that were found in BLG[14-17] and TLG[18].

r-FLG sheets are obtained by mechanical exfoliation onto degenerately doped silicon wafers with a 300 nm $SiO_2$ surface oxide layer, and identified by color contrast under an optical microscope and Raman spectroscopy[11,12]. A multi-level electron beam lithography process defines electrodes and a suspended top gate[38, 39], followed by submerging the entire device in hydrofluoric acid to partially remove the underlying $SiO_2$ layer. The dual-gated geometry allows to control charge density $n$ and out-of-plane electric field $E_\perp$ independently. Typical devices have a source-drain electrode separation of 1-2 μm, and are designed have an aspect ratio of unity, though the actual aspect ratio may vary from 0.5 to 1.5 (Fig. 1a inset).

As-fabricated suspended graphene samples have poor mobility due to contamination from the electron beam resist, which can be removed through Joule heating by current annealing at low temperature[40]. Current annealing is performed by slowly and repeatedly ramping up and down the source-drain voltage bias $V$ applied to the sample, each time reaching a slightly higher maximum bias (Fig. 1a). In Fig. 1b, the two-terminal differential conductance $G=dI/dV$ of an r-4LG device is plotted as a function of back gate voltage $V_{bg}$ after repeated annealing cycles. The red, orange, green, blue and purple traces correspond to the first, fourth, fifth, seventh and eighth annealing cycles, respectively, illustrating progressively higher electron mobility, which increases by 2 orders of magnitude from 2800 to 115,000 $cm^2/Vs$, while the corresponding minimum conductance decreases from 870 to 33 μS. The maximum current density is typically ~ 0.2-0.4 mA per μm per layer.

After successful current annealing, the devices are measured in a $He^3$ or pumped $He^4$ refrigerator using standard lock-in techniques. We first examine the device in the absence of external fields. Fig. 2a presents the two-dimensional plot of differential conductance $G$ vs source-drain bias $V$ and charge density $n$ at $B=E_\perp=0$ for device 1. The transfer curve $G(n)$ at zero bias is characteristically V-shaped, similar to that observed in other graphene systems (Fig. 2b). At the charge neutrality point (CNP), the device is insulating, with conductance <0.1 μS at small bias. As source-drain bias $V$ is raised, $G$ remains negligible until $V$ reaches ±77 mV, at which point $G$ rises to extremely sharp peaks. For $|V|>90$ mV, $G$ drops to ~ 180 μS (Fig. 2c). Such a $G(V)$ characteristic in charge neutral r-4LG strongly resembles the density of states of a gapped phase, with an energy gap that corresponds to half of the peak-to-peak separation, $\Delta$~80 meV. Similar results were observed in several other r-4LG devices, with gap sizes varying from 48 to 80 meV. As the device becomes slightly electron- or hole-doped, such gap-like features disappear quickly.

At high densities, the *G(V)* curves become mostly flat, indicating that the device reverts to a conventional metal.

The insulating gapped state in charge neutral r-4LG is unexpected in single particle band structure calculations. In principle, a band gap may appear if the device has a built-in interlayer potential that breaks the inversion symmetry, *e.g.* when the top and bottom layers have impurities with equal and opposite charges, so that overall the device remains charge neutral but hosts a potential difference between the outmost layers. In this case, the gap should be enhanced (reduced) if an external out-of-plane electric field $E_\perp$ is applied in the same (opposite) direction of the pre-existing interlayer potential. Alternatively, this gapped state may be a ground state with spontaneous broken symmetries, with an intrinsic gap arising from strong electronic interactions due to the extremely flat bands and large density of states at the charge neutrality point. Such an interaction-induced gap should have symmetric response to $E_\perp$ of either polarity.

To ascertain the nature of the gapped state observed in r-4LG, we examine its dependence on an out-of-plane electric field $E_\perp$. Fig. 3a plots $G(V, E_\perp)$ at *n=B=0* for another device, where the dark blue region corresponds to the insulating state, and the red "lips" to the sharp peaks in *G*, indicating a gap $\Delta$~50 meV at $E_\perp$=0. With increasing $|E_\perp|$, $\Delta$ decreases (Fig. 3b), while accompanied by slowly increasing zero bias conductance (Fig. 3c). At $E_\perp = \pm 156$ mV/nm, the gap completely closes, and *G* reaches a maximum at ~4 $e^2/h$, with flat *G(V)* curves. Interestingly, the dark blue region, or a gapped state, re-appears when $|E_\perp|$ increases further, as *G* decreases and *G(V)* once again exhibits a valley near zero bias.

Such symmetric and non-monotonic dependence of *G(V)* on $E_\perp$ establishes that the gap observed at the charge neutrality point is not the result of the single-particle state with built-in interlayer potential, but rather emerges from electronic interactions. The initial closing and subsequent re-opening of the gap by $|E_\perp|$ is particularly interesting, as it suggests the presence of two distinct gapped states with different symmetries in the bulk. At sufficiently large $|E_\perp|$, most charges are expected to reside on the topmost or bottommost layer, thus we identify the re-appeared gapped state at $|E_\perp|>150$ mV/nm as a layer polarized insulator. The intrinsic insulating state at $E_\perp=B=n=0$ is consistent with a layer antiferromagnetic state, in which the spins in the top and bottom surface layers are oppositely spin polarized, similar to that observed in BLG[14-17], r-TLG[18, 19], and ABA-stacked FLG with even number of layers[41, 42].

To quantitatively investigate dependence of $\Delta$ on interlayer bias, we note that the applied $E_\perp$ as imposed by external gates is heavily screened. Thus the actual interlayer potential difference *U* between the top and bottom layers is much smaller than the applied external potential $U_{ext}$= $E_\perp d$, where *d*=1 nm is the thickness of 4LG. To calculate *U* for a given $U_{ext}$, we use a simplified two-band model, where only nearest intralayer hopping $\gamma_0$ and interlayer hopping $\gamma_1$ are considered. The Hamiltonian

$$H = \begin{pmatrix} -\frac{U}{2} & \frac{v^4}{\gamma_1^3}(\xi p_x - ip_y)^4 \\ \frac{v^4}{\gamma_1^3}(\xi p_x + ip_y)^4 & \frac{U}{2} \end{pmatrix}$$

acts on the wave function $\psi = \begin{pmatrix} \psi_{A1} \\ \psi_{B4} \end{pmatrix}$, where $v = \frac{\sqrt{3}a\gamma_0}{2\hbar} \sim 10^6$ m/s is the Fermi velocity, a = 2.46 Å is the distance between unit cells, $\xi = +1$ (-1) for the $K_+$ ($K_-$) valley, $p$ is the momentum, and A1 and B4 denote the A-sublattice from the bottommost layer and B-sublattice from the topmost layer, respectively. The energy eigenvalue for the equation $H\psi = E\psi$ is $E = \pm\sqrt{\frac{U^2}{4} + \frac{v^8 p^8}{\gamma_1^6}}$, with wave function

$$\psi = \sqrt{\frac{E - \frac{U}{2}}{2E}} \left( -\frac{v^4 p^4}{\gamma_1^3} \frac{1}{\left(E - \frac{U}{2}\right)} e^{3i\varphi\xi} \right) e^{i\vec{p}\cdot\vec{r}/\hbar}$$

where $\varphi = tan^{-1}(p_x/p_y)$. From electrostatics, we can write

$$U = U_{ext} + \frac{de^2}{2\varepsilon_o \varepsilon_r}(n_4 - n_1)$$

where $\varepsilon_0$ is the permittivity of vacuum, and $\varepsilon_r$ is the dielectric constant of 4LG. $n_1$ and $n_4$ are the charge density of the bottom and top layers, respectively, given by

$$n_1 = \frac{2}{\pi\hbar^2}\int|\psi_{A1}|^2 p dp, \qquad n_4 = \frac{2}{\pi\hbar^2}\int|\psi_{B4}|^2 p dp$$

At the charge neutrality point, screening is provided by the filled valence band, yielding

$$U = U_{ext} - \frac{de^2}{2\varepsilon_o\varepsilon_r}\left(\frac{1}{\pi}\right)\left(\frac{\gamma_1}{\hbar v}\right)^2 \left(\frac{U}{2\gamma_1}\right)^{1/2} \left[\int_0^\infty \frac{dx}{\sqrt{1+x^4}}\right] \qquad (1)$$

which can be solved numerically to yield $U$ for a given value of applied $U_{ext}$ (Fig. 3c inset, assuming $\varepsilon_r = 1$).

Using Eq. (1), we calculate $U$ for applied $E_\perp$, and plot the extracted gap $\Delta$ from Fig. 3a as a function of $U$ (Fig. 3d). After correcting for screening, $\Delta$ is suppressed linearly by the interlayer potential, with a slope of 6.5 meV/mV.

To further elucidate the nature of the interaction-induced gap in r-4LG, we investigate its dependence on an out-of-plane magnetic field $B$. Fig. 4a-b plots $G(V, B)$ at $n=E_\perp=0$, and line traces $G(V)$ at $B$= 0, 0.5, 1, 1.5, and 2 T, respectively. As $B$ increases, the gap is widened, again consistent with observations in BLG[15]. Fig. 4c illustrates that $\Delta$ scales approximately linearly with $B$ for three different devices, albeit with different slopes, presumably due to variations in sample quality. On the other hand, the heights of resistance peaks *decrease* with increasing $B$, a trend that is opposite to that in BLG [15]. This suggests that at large $B$, the gap is no longer a "hard gap". We currently do not understand this phenomenon, which will be investigated in future studies.

Lastly, we explore the temperature dependence of r-4LG devices. A representative data set of the evolution of $G(V)$ curves at $n=E_\perp=B=0$ is shown in Fig. 5a. As temperature increases, the sharp peaks in $G(V)$ move closer, and their heights decrease while the minimum conductance increases. At $T=70K$, the $G(V)$ curves are almost flat, indicating a complete closure of the gap as the device behaves as a conventional resistor. Fig. 5b plots the normalized gap $\Delta/\Delta_0$ as a function of $T$ for three different samples, where $\Delta_0$ is the gap measured at the lowest temperature (0.3 – 1.5K). All data fall on the same curve, indicating great consistency among the samples. The data is fit to the function[26]

$$\Delta(T) = \Delta(0)\left[A\left(1-\frac{T}{T_c}\right) + B\left(1-\frac{T}{T_c}\right)^2\right]^{1/2} \qquad (2)$$

where $T_c$ is the critical temperature. Eq. (2) reduces to the usual mean–field functional form for a phase transition $\sqrt{1-\frac{T}{T_c}}$ for $T/T_c \rightarrow 1$, and the second term $\left(1-\frac{T}{T_c}\right)^2$ is inserted to capture the vanishingly small dependence on $T$ as it approaches 0. Excellent agreement with data is obtained, yielding $T_c$=40.4K.

Fig. 5c plots the measured minimum conductance $G_{min}(T)$ data for the three devices. All three devices exhibit very similar behavior – $G_{min}$ scales approximately linearly with $T$ at high temperatures ($T$>60K), but drops precipitously to ~0 at $T$<30 K, as the devices undergo a metal-insulator transition. The apparently large variations in magnitude of $G_{min}$ may be accounted for by the different aspect ratios of the devices. Indeed, when normalizing $G_{min}$ by $G_{min}(T=300K)$, all three curves collapse onto the same line, with an apparent $T_c$ of ~ 40K (Fig. 5d). Naively, in the absence of electronic interactions, the minimum conductivity $\sigma_{min}$ of an $M$-layer graphene is expected to be on the order of $4Me^2/\pi h$; thus, for r-4LG, one expects $\sigma_{min}$~$5e^2/h$ ~ 200 μS. Extrapolating the high temperature behavior of the devices to $T$=0, we obtain $G_{min,extp}$ that range from 150 to 450 μS. As electronic interactions are minimal at high temperatures, and considering that the samples have aspect ratio ranging from 0.5 to 1.5, these values of $G_{min,extp}$ are in reasonable agreement of $\sigma_{min}$ from band structure calculations.

In summary, we observe an interaction-induced insulating state in charge neutral r-4LG in the absence of external fields, with an energy gap as large as 80 meV. This gap is further enhanced by the application of an out-of-plane magnetic field, and closed by increasing temperature, charge density and an out-of-plane electric field of either polarity. It is similar to that observed in BLG and r-TLG, and therefore identified to be a layer antiferromagnetic state with broken time reversal and spin rotation symmetries. The large magnitude of the gap also suggests that, at least in tetralayer graphene, the band flattening effect "wins" over the increasing screening and charge de-confinement in thicker samples, though the rate of increase in gap size with the addition of another atomic layer is smaller (see Table 1). Whether the gap will continue to increase and the nature of the ground state in rhombohedra-stacked pentalayer and thicker graphene warrant future studies.

Acknowledgments: We thank the cleanroom staff at the Center for Nanoscale Science and Engineering in the Bourns College of Engineering at UC Riverside. The experiments are supported by DOE BES Division under grant no. ER 46940-DE-SC0010597. Part of this work was performed at the National High Magnetic Field Laboratory that is supported by NSF/DMR-0654118, the State of Florida, and DOE.

**Table 1.** Electron energy as a function of wave vector $k$, carrier density $n$, density of states (DOS), interaction parameter $r_s$, measured gap size $\Delta$ and measured critical temperature $T_c$ for SLG, BLG, r-TLG and r-4LG.

|  | $E(k)$ | $E(n)$ | DOS | $r_s$ | $\Delta$ (meV) | $T_c$ (K) |
|---|---|---|---|---|---|---|
| MLG | $\propto k$ | $\propto n^{1/2}$ | $\propto n^{1/2}$ | constant | <0.1 meV |  |
| BLG (ref. 14-17) | $\propto k^2$ | $\propto n$ | *constant* | $\propto n^{-1/2}$ | 2 | 5 |
| r-TLG (ref. 18) | $\propto k^3$ | $\propto n^{3/2}$ | $\propto n^{-1/2}$ | $\propto n^{-1}$ | 42 | 36 |
| r-4LG (this work) | $\propto k^4$ | $\propto n^2$ | $\propto n^{-1}$ | $\propto n^{-3/2}$ | 80 | 40 |

**Fig. 1.** Current annealing and transport characterization of r-4LG devices. **a.** *I(V)* curves during current annealing. The red and purple curves correspond to the first and eighth annealing cycles. **Inset.** False color SEM image of a dual-gated suspended FLG device. **b.** $G(V_{bg})$ after different annealing cycles. Red, orange, green, blue and purple lines correspond to the first, fourth, fifth, seventh and eighth annealing cycles, respectively.

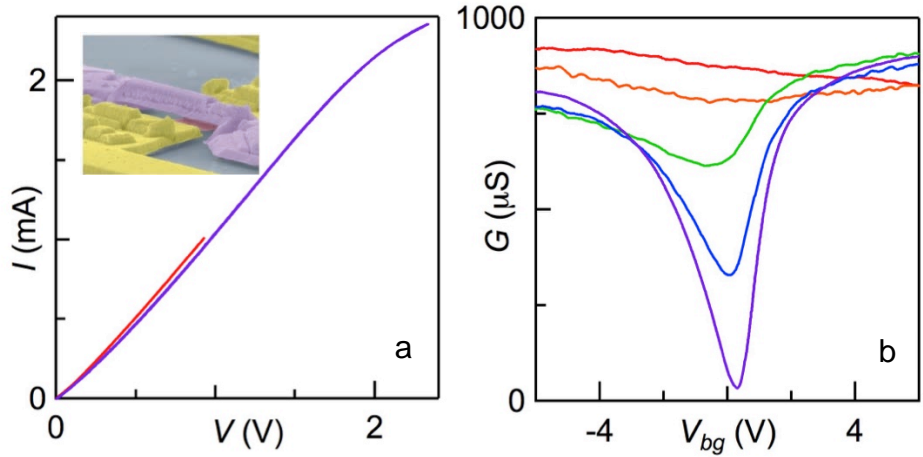

**Fig. 2.** Charge density *n* dependent transport spectroscopy at $B=E_\perp=0$. **a.** 2-termianl differential conductance $G(V, n)$ in mS. **b.** $G(n)$ line cut at $V=0$. **c.** $G(V)$ line cut at different charge densities.

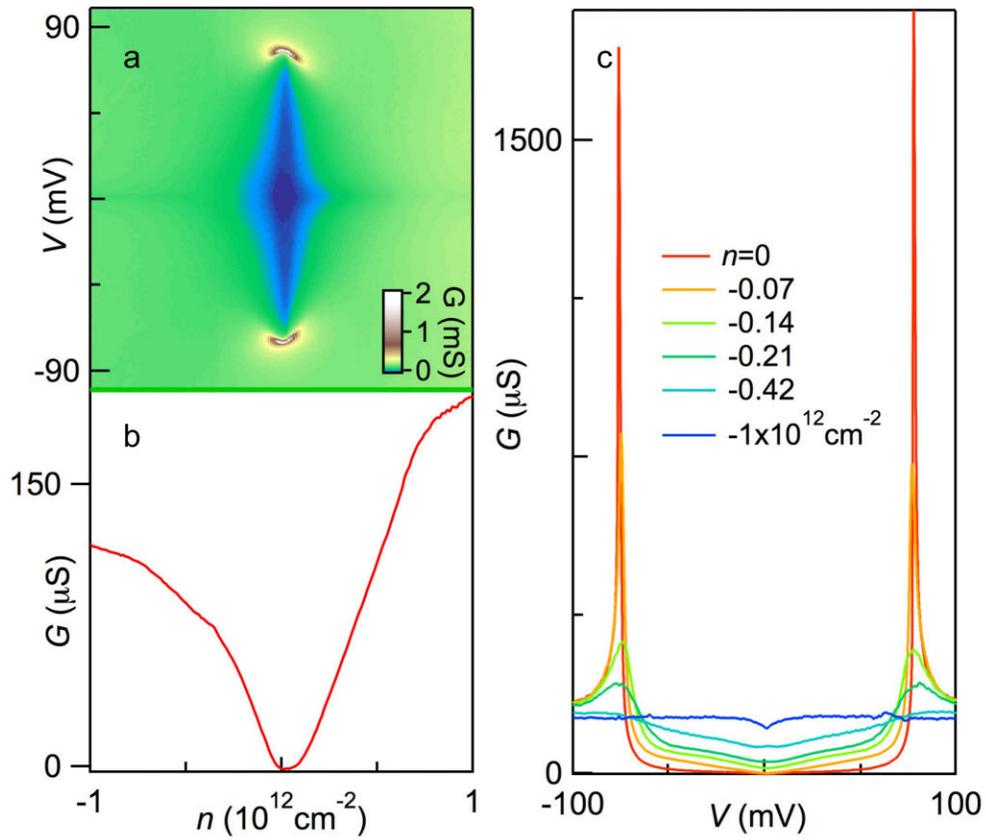

**Fig. 3.** $E_\perp$-dependent transport spectroscopy at $B=n=0$. **a.** Two-termianl differential conductance $G(V, E_\perp)$. **b.** $G(V)$ line cuts at $E_\perp=0$ (blue), 75 (green), 155 (orange) and 190 (red) mV/nm, respectively. **c.** $G(E_\perp)$ line cut at $V=0$. Inset. **Inset:** Actual interlayer potential difference $U$ as a function of applied external potential $U_{ext}$. **d.** Gap size $\Delta$ as a function of $U$ after correction for screening.

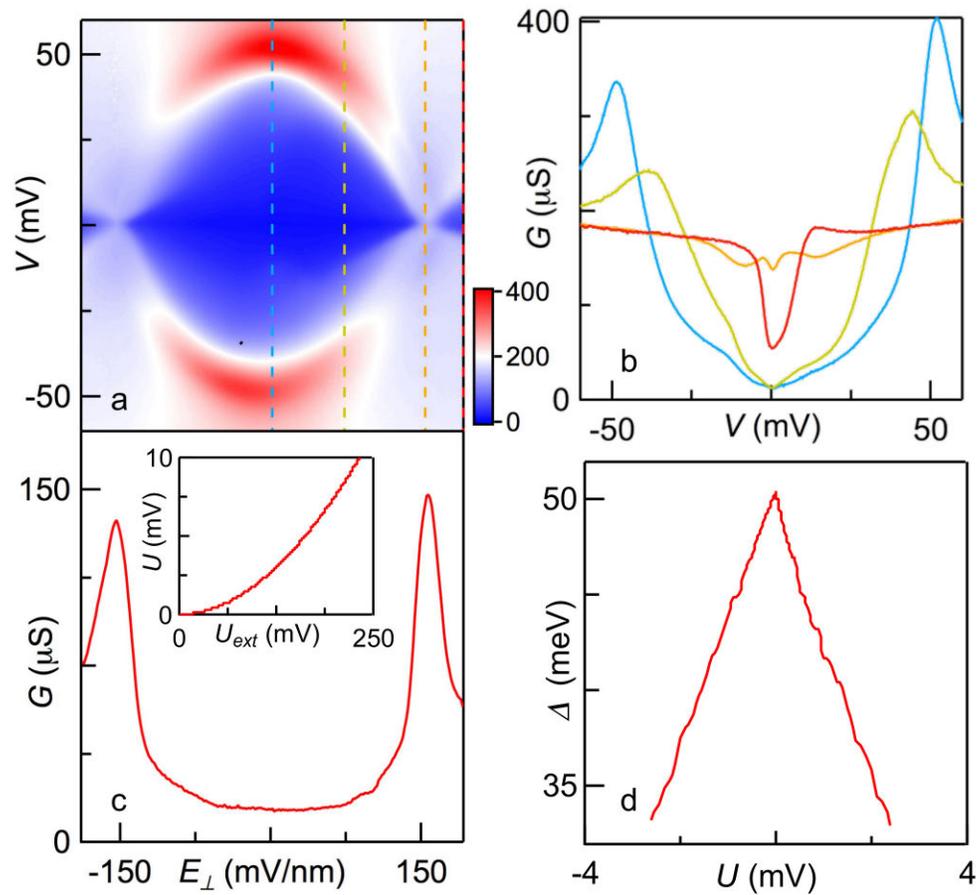

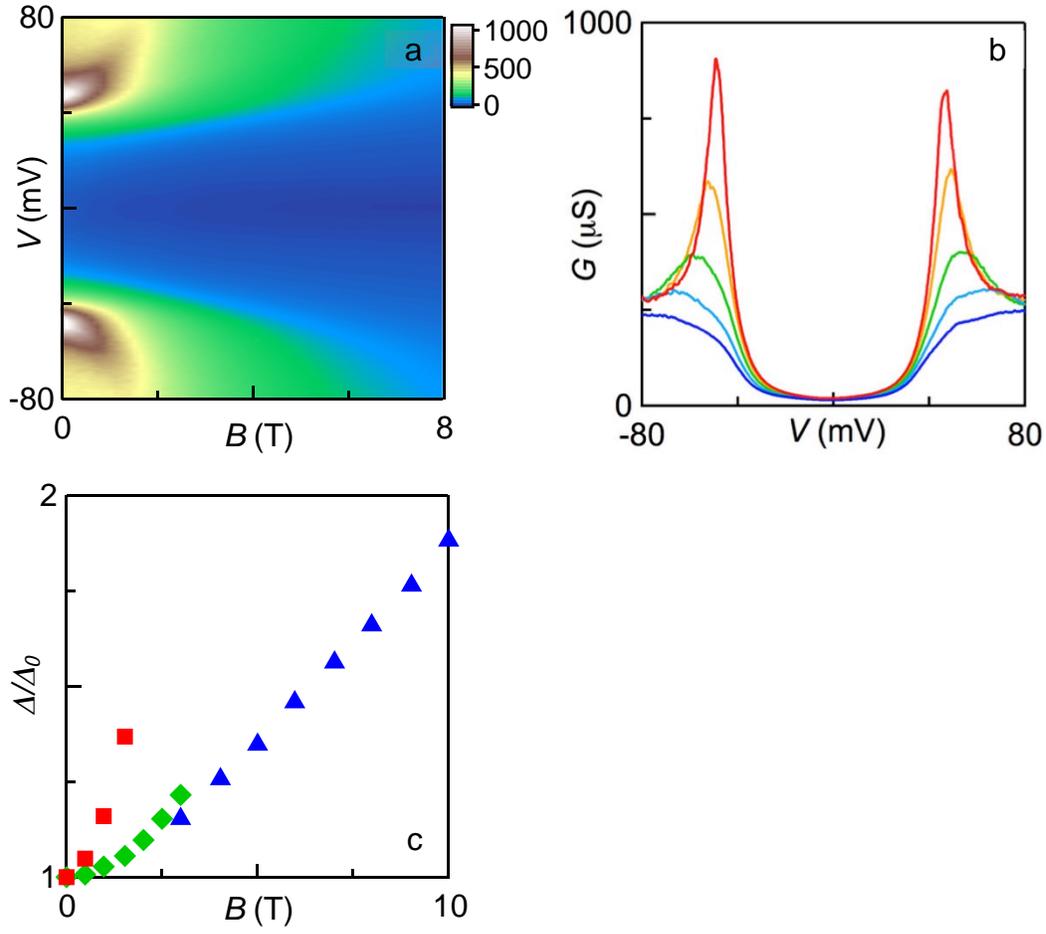

**Fig. 4.** Transport data at $n=E_\perp=0$ at different magnetic fields. **a.** $G(V,B)$ in μS. **b.** Line traces $G(V)$ from (a) at $B=0$ (red), 0.5 (orange), 1 (green), 1.5 (light blue), and 2T (dark blue), respectively. **c.** Normalized gap $\Delta/\Delta(0)$ from three different devices plotted versus $B$.

**Fig. 5.** Transport spectroscopy at $n= E_\perp=B=0$ at different temperatures. **a.** $G(V)$ at 70 (red), 35 (orange), 30 (green), 20 (light blue), and 0.260 K (dark blue), respectively. **b.** Normalized gap $\Delta(T)/\Delta_0(0)$ for three different devices. **c-d.** $G_{min}(T)$ and normalized minimum conductance $G_{min}(T)/G_{min}(300K)$ for these devices.

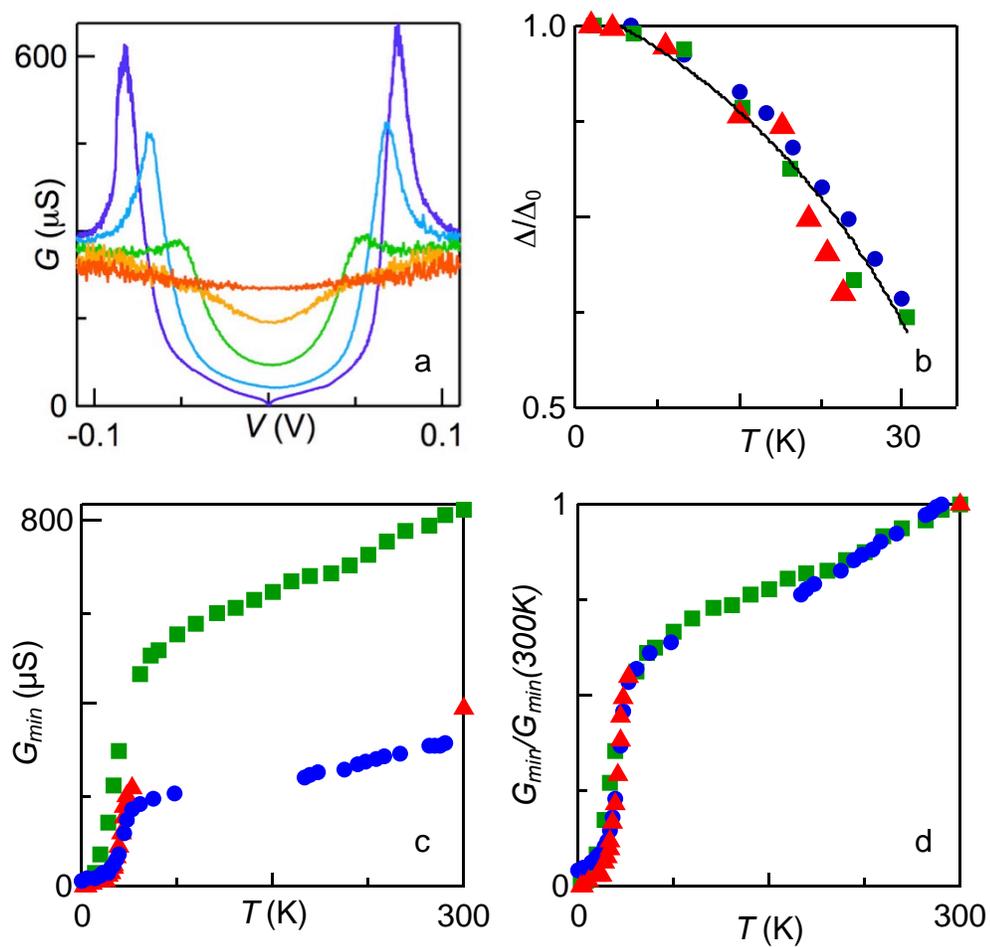